\documentclass[a4paper,11pt]{article}
\usepackage{pos}

\newcommand{\bea}{\begin{eqnarray}}
\newcommand{\eea}{\end{eqnarray}}

\title{Real time simulations of scalar fields with kernelled complex Langevin equation}

\author[a]{Daniel Alvestad}
\author[a]{Alexander Rothkopf}
\author*[b]{D\'enes Sexty}

\affiliation[a]{
Faculty of Science and Technology, University of Stavanger, \\ 4021 Stavanger, Norway
}

\affiliation[b]{
Institute of Physics, NAWI Graz, University of Graz,\\
Universit\"atsplatz 5, 8010 Graz, Austria}

\emailAdd{alexander.rothkopf@uis.no}
\emailAdd{denes.sexty@uni-graz.at}

\abstract{ Real time evolution of a scalar field theory is investigated. The severe sign problem is circumvented
using the Complex Langevin equation. The naive application of the method breaks down for extended real times due to the appearance of boundary terms. We use the kernel freedom of the complex Langevin equation to push the breakdown to larger real-times. We search for the optimal kernel using machine learning methods. Thus, we extend the available range for 1+1d scalar simulations beyond the state of the art simulations.}

\FullConference{The 41st International Symposium on Lattice Field Theory (LATTICE2024)\\
 28 July - 3 August 2024\\
Liverpool, UK\\}

\begin{document}
\maketitle

\section{Introduction}

Understanding the dynamics of quantum systems
requires the ability to calculate real-time
correlators in such systems. This applies in 
many areas in theoretical physics,
be it condensed matter physics, nuclear-, particle- physics or cosmology.
However, such calculations 
currently rely 
on various approximation
schemes, as a genuine non-perturbative ab initio approach remains 
unavailable.
Lattice discretisation
is a very successful ab initio approach to equilibrium
physics (and static correlators), using 
the Euclidean-time formalism. In contrast, in the 
real-time formalism, the lattice has a very 
severe sign problem, hampering the usual 
Markov-chain Monte Carlo methods. This is a 
consequence of the fact that in the path-integral formalism,
the weight of field configurations for real times is a phase factor.

Many approaches have been invented to circumvent 
the sign problem on the lattice, among others 
dual variables \cite{gattringer_approaches_2016} 
which do solve the sign problem in certain cases,
however their applicability is limited,
extrapolation \cite{de_forcrand_constraining_2010,braun_imaginary_2013,braun_zero-temperature_2015,guenther_qcd_2017} and reweighting, 
which work as long as the sign problem is relatively mild, and there is a nearby ensemble with positive
measure to use as a 'stand in' for the ensemble 
to be studied,
and finally, methods using the analytic properties
of the theory: complex Langevin \cite{Klauder:1983nn,Parisi:1984cs} 
and Lefschetz thimbles and related methods \cite{rom_shifted-contour_1997,cristoforetti_new_2012,Alexandru:2020wrj,Nishimura:2023dky}.

Quantum computers offer a different way to 
tackle the sign problem by e.g. building 
an analog quantum system in the quantum computer, such that unitary time evolution is realized. 
Currently interesting systems are out of the 
reach of quantum computers, however this might change in the future \cite{bauer_quantum_2022,Dalzell:2023ywa}.

Here we use the Complex Langevin equation (CLE)
to tackle the sign problem.
We study the time evolution of 
a scalar field with quartic interactions in 1+1 dimensions using 
the complex Langevin equation (CLE) \cite{Alvestad:2023jgl}.
A study of the same system using a 
Lefschetz-thimble inspired optimized manifold has 
appeared in \cite{Alexandru:2017lqr}, 
which we take as a benchmark calculation to 
assess the performance of the CL setup.
The real-time scalar system in 0+1 dimensions as well as pure-gauge in 3+1 dimensions have been studied before using the CLE in \cite{Berges:2006xc,Berges:2007nr}.
Recently it has been realized that using the 
Kernel freedom of the complex Langevin equation
(which is similar to the introduction
of a field dependent diffusion constant for 
the case of the real Langevin equation),
one can influence the boundary terms 
and simulations for longer real-time extents are 
possible \cite{Alvestad:2022abf,Lampl:2023xpb}.
To achieve this, one optimizes the kernel using 
a suitable loss function.
Here we extend these studies to 1+1 dimensions,
using the optimization procedure developed in the above mentioned two studies. in particular we deploy the highly efficient loss function proposed in \cite{Lampl:2023xpb}.

In section 2, we shortly introduce the Complex Langevin equation with a kernel, 
in section 3 we describe our theory and our numerical setup,
in section 4 we show numerical results.

\section{Complex Langevin equation and Kernel}

For scalar fields $x_i$, the Langevin equation reads as
\bea
dx_i  = - { \partial S \over \partial x_i } d \tau + 2 d w_i
\eea
where $ dw $ is the increment of a standard Wiener process with $ \langle d w_i = 0 \rangle, \ \langle dw_i dw_j \rangle =  2 \delta_{ij} d \tau $, and $S(x_i)$ is the action.

The Langevin equation can be modified with a matrix $H_{jk} $, which can be 
an arbitrary function of the scalar fields. 
\bea \label{kernelledlang}
dx_i  = -  H_{ji} H_{jk} { \partial S \over \partial x_k } d \tau + {\partial  H_{ji} H_{jk }  \over \partial x_k     } d\tau + 2 H_{ij} d w_j
\eea
One usually calls the matrix $ K = H^T H$ the kernel.
For the real Langevin equation, under rather mild assumptions, the stationary solution of the Fokker-Planck equation 
 is unchanged under the introduction of a kernel.
(The Fokker-Planck equation governs the Langevin-time evolution of the probability distribution $ P(x_i,\tau)$.)
For the CLE,
in contrast to the real case, the stationary distribution on the complexified 
manifold does change.
To ensure that the analytic continuation
is not broken, we require the kernel to be a holomorphic function.
In certain cases the introduction of the kernel can influence the appearance of the boundary terms \cite{Okamoto:1988ru,Okano:1991tz}.
One observes that a kernel can in principle also cause the CLE to deliver incorrect results \cite{Alvestad:2022abf}, recently understood to correspond to an integration cycle different than the desired real cycle
(for details see \cite{Hansen:2024kjm}). 
We use boundary terms for diagnostic purposes,
see \cite{boundaryterms1,boundaryterms2} for their definition and properties.

In this study we exploit the kernel degree of freedom to improve convergence properties, i.e. suppress
boundary terms of the CLE by optimizing a kernel with machine learning methods, to be 
described in the next section.

\section{Real-time scalar fields}

\begin{figure}
\centering
\includegraphics[scale=0.7, trim= 0cm 0.2cm 0 0.5cm, clip=true]{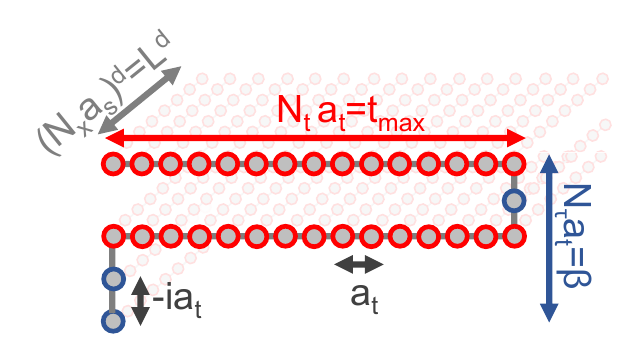}
\caption{Geometry of discretized (d+1) dimensional scalar field theory on the Schwinger-Keldysh contour. Here we use $N_t=32$, $N_\tau=4$ and $N_x=8$ and $a_tm=1/10, a_sm=2/10$.}\label{fig:SKTthermdiscr}
\vspace{-0.3cm}
\end{figure}

We investigate a scalar field theory with quartic coupling,
described by the action, written in Minkowski metric as
\bea
S = \int d^dx dt {1\over 2 } \left( (\partial_t \phi)^2 - (\partial_i x )^2  - m^2 \phi^2 - {\lambda \over 12 } \phi^4 \right).
\eea
To describe temporal correlators such as $ \langle \phi(t) \phi(0) \rangle$ we write in Heisenberg picture
\bea
\langle \phi(t) \phi(0) \rangle = \langle e^{i\hat Ht}  \phi  e^{-i\hat Ht} \phi \rangle
\eea
with $\hat H$ the Hamiltonian of the system.
In path-integral formalism, to describe this correlator, we need to define
our theory on a Schwinger-Keldish contour, 
going first from $t=0$ to some $t_\textrm{max} \ge t $, then turning back and continuing from $ t_\textrm{max}$ to $ t=0$, 
implementing the two time-evolution operators in the above formula.
We can also describe thermal averages by using an imaginary time path at the end of the path so far, going from $t=0$ to $ t=-i \beta$ 
with the inverse temperature $\beta$,
as well as choosing periodical boundary conditions on the fields: $ \phi(t=0)=\phi(-i \beta)$, implementing
\bea
 \textrm{Tr}( e^{-\beta \hat H} \phi(t) \phi(0) ) = \textrm{Tr} ( e^{-\beta \hat H}  e^{i\hat Ht} \phi(t) e^{-i\hat H t}  \phi(0) ).
\eea
If we are interested in time ordered correlators, and since we have a time-independent Hamiltonian
the second and third part of the contour can be replaced with any time-path starting at $t_\textrm{max}$ and 
ending at $ - i \beta$. In practice we use a contour such that we split the vertical path in to two
halves, one between the forward and backward branches and one at the end, as depicted in Fig.~\ref{fig:SKTthermdiscr}.
We discretize the 1+1 dimensional spacetime using the stepsize $ a_s m =0.2$ in spatial direction and $ a_t m =0.1 $ in temporal direction.
We discretise the action to yield 
\bea
S &=& \sum_{n,k} a_s a_{t,n} \left[     
 {1\over 2 } \left( \phi_{n+1,k} - \phi_{n,k} \over a_{t,n} \right)^2 
 +{1\over 4 } 
 \left(  
\left( \phi_{n,k+1} - \phi_{n,k} \over a_s \right)^2 +
\left( \phi_{n+1,k+1} - \phi_{n+1,k} \over a_s \right)^2
\right) \right. \\ \nonumber
&& \left.+{1\over 4} m^2 \left( \phi_{n,k}^2 + \phi_{n+1,k}^2 
\right)
+ {1\over 2 \cdot 4! }\left( \phi_{n,k}^4 + \phi_{n+1,k}^4 
\right)
 \right],
\eea
where $ \phi_{n,k} $ is the scalar field on the discretised spacetime at the $n$-th time slice and 
$k$-th spatial point, and $ \gamma_n $ describes the discretised complex time-path, such that 
$ a_{t,n} = \gamma_{n+1} - \gamma_n$.

To simulate the theory we use the kernelled complex Langevin equation (\ref{kernelledlang}), with a 
field independent matrix kernel. To optimize the matrix, we use Machine Learning inspired techniques, with 
the loss function (for more discussion see \cite{Alvestad:2023jgl})
\bea
 L = \int d^d x dt (\textrm{Im} \phi(x,t))^2.
\eea

\section{Results}

For numerical simulations we used $ \lambda =1 $ and 
$ \beta m = 0.4$ as also used in \cite{Alexandru:2017lqr}.
The lattice consisted of $ N_t=32$ points 
along the forward and backward branches of the contour and 
$ N_\tau=4$ along the vertical branch. We used $N_s=8$ points along the spatial direction.
Using temporal stepsize $ a_tm =0.1$ thus we have $t_\textrm{max}=3.2$, such 
that naive complex Langevin (i.e. using $H=1$) breaks down 
on this contour due to large fluctuations and large boundary terms.
We used an adaptively controlled Langevin update with maximal stepsize $ \Delta \tau_\textrm{max}=0.001 $.

Our observables are the equal time correlation function
\bea \label{equaltimecorrelator}
F(\gamma) = \langle \phi(\gamma,p=0) \phi( \gamma, p=0) \rangle
\eea
and the unequal time two point function
\bea \label{unequalcorrelator}
C(\gamma) = \langle \phi(\gamma,p=0) \phi( 0, p=0) \rangle.
\eea
Since the system is in thermal equilibrium, $ F(t) $ is independent
of the time, and it can also be measured on an imaginary 
time contour, $C(t)$ shows some nontrivial 
temporal dependence.
Note that we have taken the zero momentum component 
in spatial directions, as this makes the correlator 
less susceptible to finite-size effects.
In 0+1 dimensions one can easily calculate $ F(t)$ and $C(t)$ using the Schr\"odinger equation \cite{Berges:2006xc}, however introducing spatial
directions makes that calculation very quickly infeasible.

The complex dense kernel is parametrized as $ H = A+iB $ with two real matrices,
which thus has $ 2 ((2 N_t + N_\tau) N_s)^2 $ parameters, where $N_s$ is the number
of lattce points in spatial direction, $N_t$ is the number of time slices on the forward and 
backward branches, and $N_\tau$ on the vertical branch of the Schwinger-Keldysh contour.
The gradients of the loss function are calculated by considering update steps using 
the current value of the kernel. In principle one update step suffices for this, but 
the autodiff capability of modern programming languages makes it easy to use 
many steps before taking the gradient. The optimization of $H$ starts from 
a unit matrix, and we use a learning rate of $ 10^{-3}$ and the Adam optimizer.
We observe a decay of the loss function from the $ L  \approx 1000$ initial 
value to $ L \approx 7 $ at the end of the optimization.

\begin{figure}
\centering
\includegraphics[scale=0.47, trim= 0cm 0.2cm 0 0.5cm, clip=true]{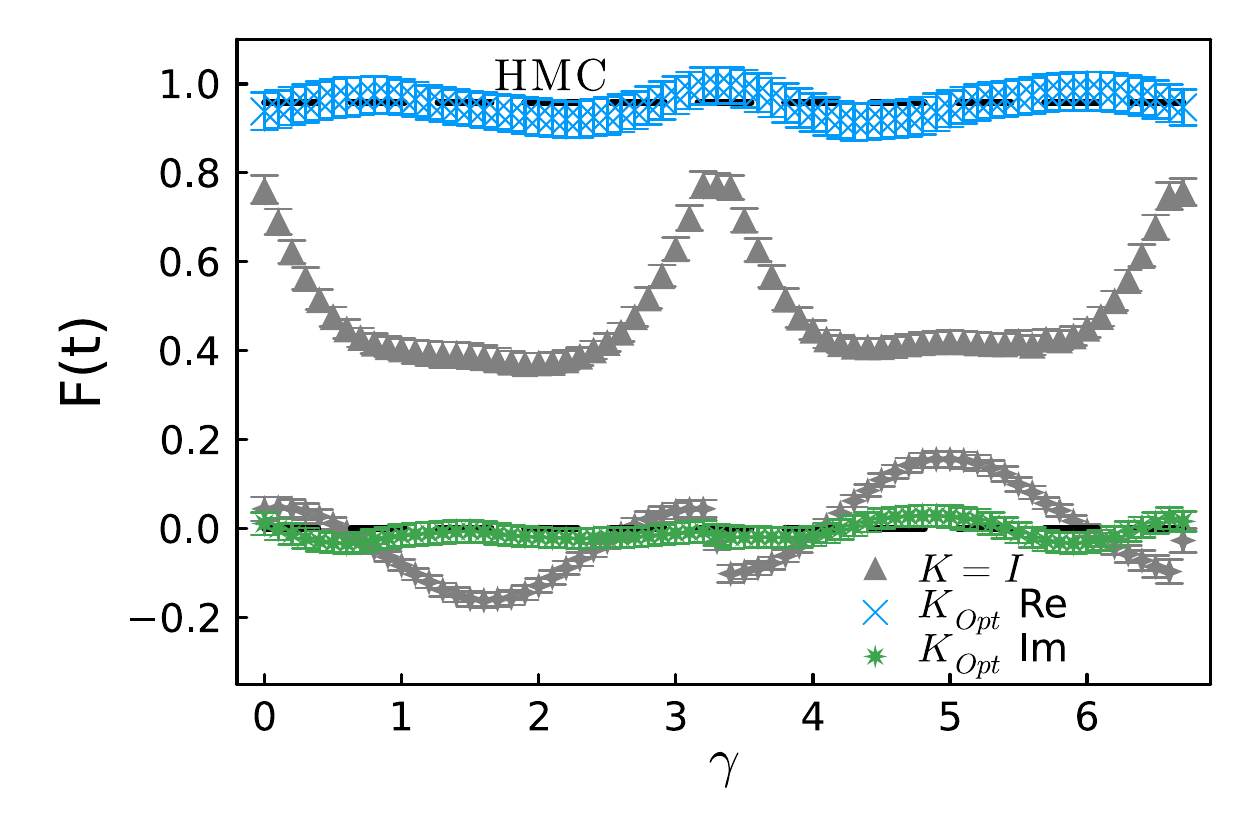}
\caption{The equal time correlator defined in (\ref{equaltimecorrelator}) as a function of the $\gamma$ parameter indexing the points on the Schwinger-Keldysh-like contour. "HMC" shows the value of the correlator as calculated on an Euclidean time path.
Grey symbols show naive CLE and and blue and green symbols show results from CLE with the optimal kernel, as indicated.  
}\label{fig:equaltime}
\vspace{-0.3cm}
\end{figure}
\begin{figure}
\centering
\includegraphics[scale=0.52, trim= 0cm 0.2cm 0 0.5cm, clip=true]{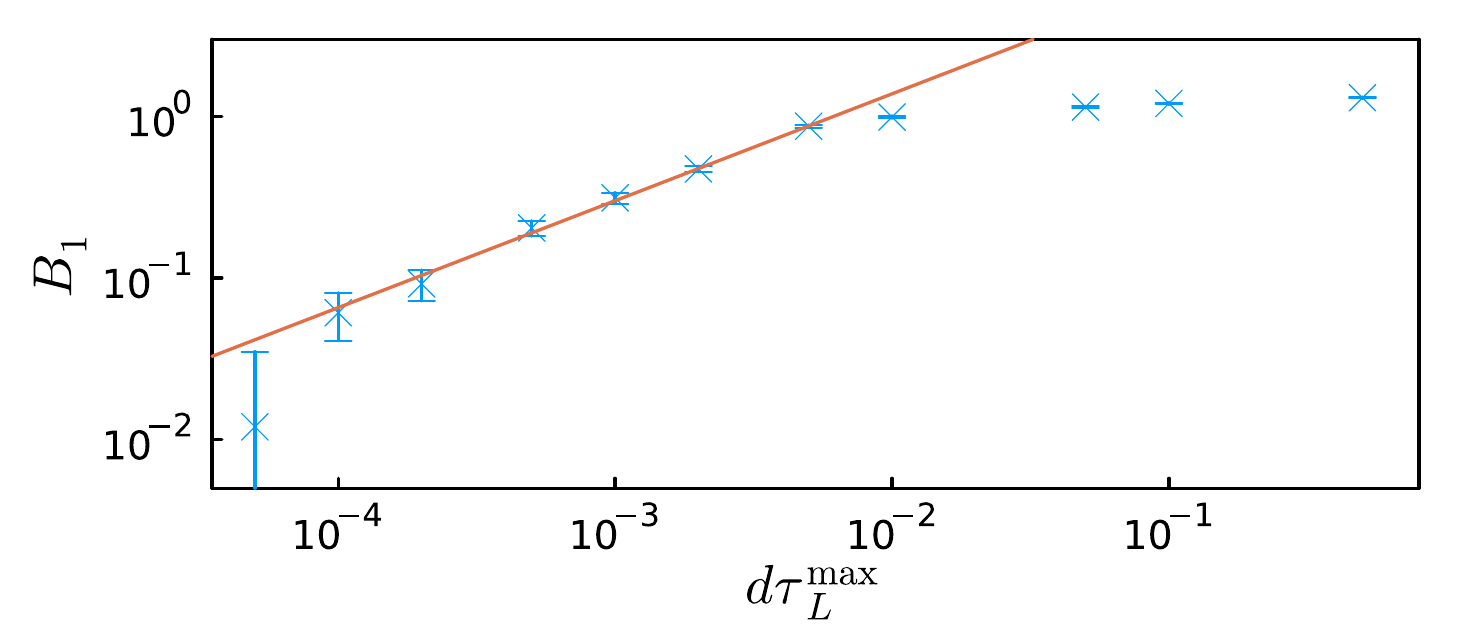}
\caption{The boundary terms of $\phi^2$ 
calculated in a CLE simulation 
with the optimized kernel, as a function 
of the maximal allowed stepsize in the adaptive 
stepsize algorithm.}
\label{fig:BT}
\vspace{-0.3cm}
\end{figure}

In Fig.~\ref{fig:equaltime} we show the measured equal time correlator for various setups.
First of all, "HMC" shows the value of the correlator as measured on a sign-problem free imaginary time-contour.
We also show the CLE results when using no kernel with grey triangles, and one observes that CLE gives clearly incorrect results in this case.
Finally, we show the CLE results with an optimized
kernel (blue and green symbols) and one observes 
time-independence, zero imaginary part and agreement 
with the "HMC" values, all of which are required from 
the correct solution.
\begin{figure}
\centering
\includegraphics[scale=0.47, trim= 0cm 0.2cm 0 0.5cm, clip=true]{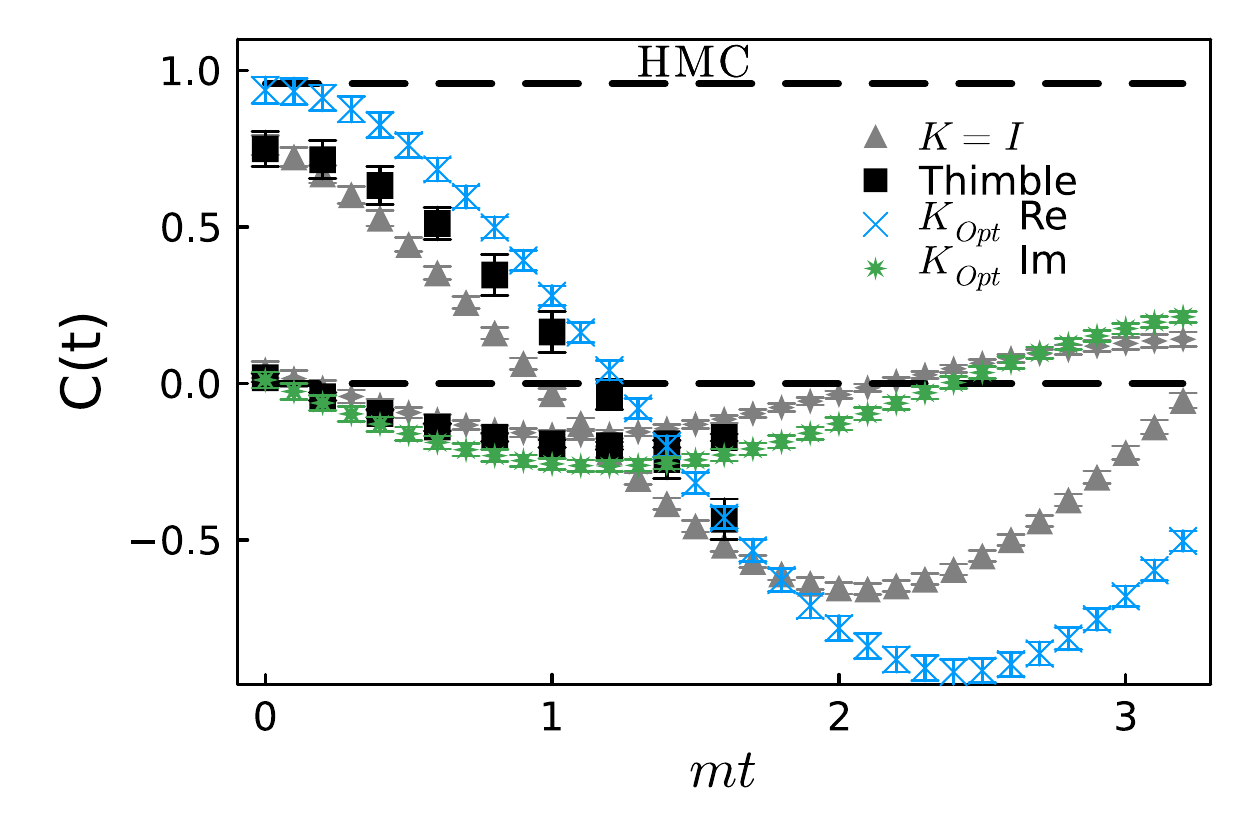}
\caption{ The unequal time correlator 
defined in (\ref{unequalcorrelator}).
The "HMC" band shows the value 
of $ C(0) = F(0) $ as measured on an imaginary time
contour.
Gray triangles show the results calculated with the 
naive CLE, blue and green symbols are calculate 
with the optimized kernel in the CLE.
We also show the results from \cite{Alexandru:2017lqr},
for an identical system except that the temporal lattice 
spacing is double of that in the CLE calculations.
}\label{fig:SKCt}
\vspace{-0.3cm}
\end{figure}
 In Fig.~\ref{fig:SKCt} we show the measured values for $C(t)$ using the naive CLE as well as the CLE with the optimized kernel, as well as the HMC
 result for $C(0)$ shown with the dashed line.
 Also here, one notes that the two CLE results are inconsistent with each other. Since we must have $ C(t=0)=F(t=0)$, 
 we conclude that the naive CLE results are incorrect.
 The thimble result \cite{Alexandru:2017lqr} is also shown with black symbols. The discrepancy at $t=0$ is likely the result of the 
 slightly different UV cutoff used in that study 
 (their $a_t$ is double of the $a_t$ we used in the CLE simulations). 
 Note that the thimble calculation was not feasible 
 for larger temporal extents due to high numerical costs.
 In contrast, the CLE calculation uses a smaller temporal stepsize and has a $t_\textrm{max}$ twice as large, as 
 our numerical costs grow in a much milder fashion when 
 the volume of spacetime is increased.
 The agreement if $C(t=0)$ suggests that our CLE results are correct. This is further corroborated by examining 
 the boundary terms of the $ \phi^2$ observable.
 The boundary terms are known to be highly sensitive to finite stepsize effects, therefore
 we examine $B_1(\phi^2)$ as a function of the 
 maximal Langevin stepsize in the adaptive control
 algorithm, as shown in Fig.~\ref{fig:BT}. 
We observe a power-law decay such that in the continuum Langevin stepsize limit no boundary terms are expected.

\begin{figure}
\centering
\includegraphics[scale=0.17, trim= 0cm 0.2cm 0 0.5cm, clip=true]{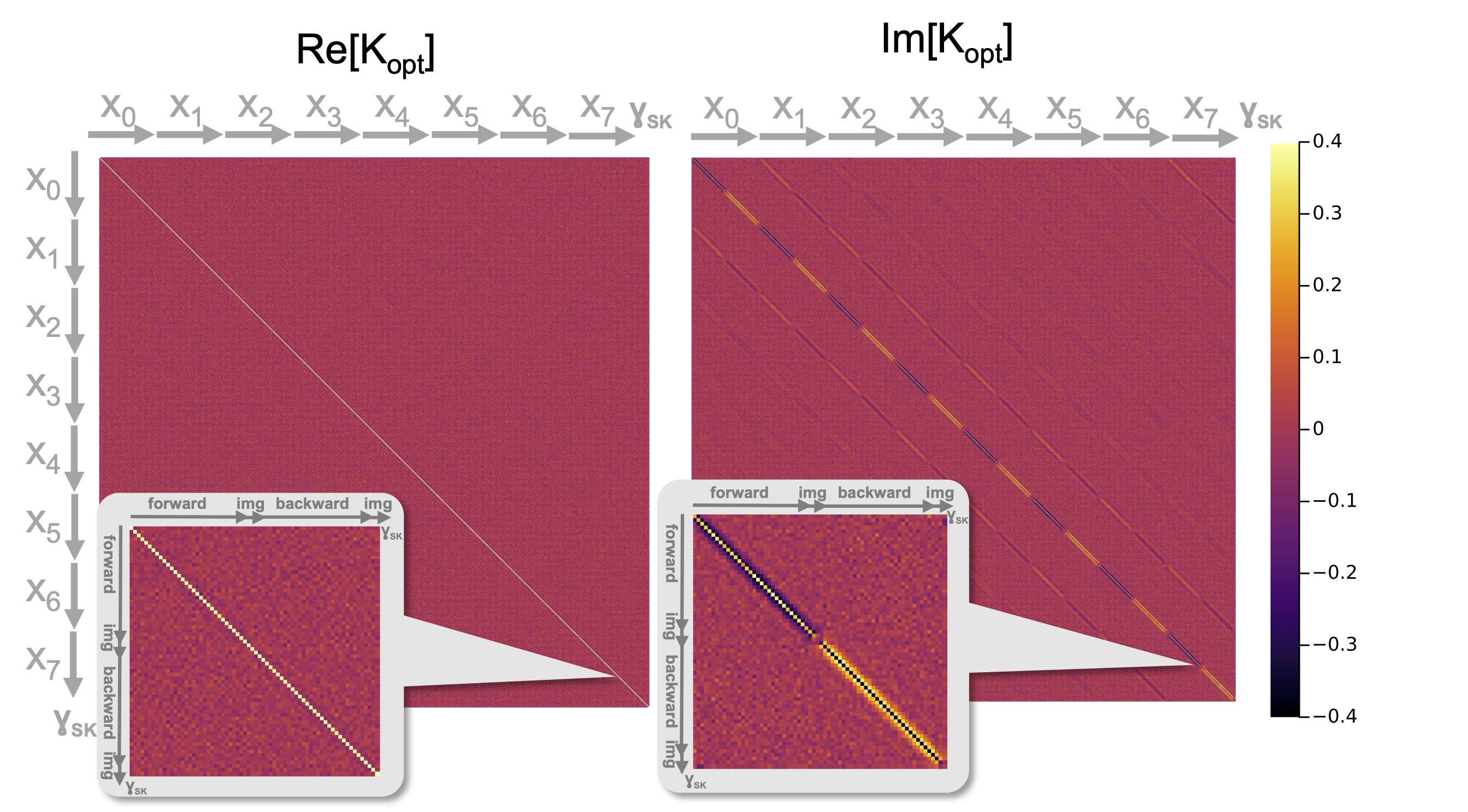}
\caption{
The heat-map of the real and imaginary 
parts of the optimized kernel of the CLE simulations. The points of the matrix are indexed such that the temporal coordinate is the fastest index.
We also show a zoom-in to a single spatial slice.
}\label{fig:kernel}
\vspace{-0.3cm}
\end{figure}
Finally, in Fig.~\ref{fig:kernel} we show the optimized kernel we used in the simulations to obtain Figs.~\ref{fig:SKTthermdiscr} and \ref{fig:SKCt}.
The color coding resolves almost all matrix values, 
except for the diagonal of the real part of the 
matrix, which is given by $\textrm{Re}[K_\textrm{opt}=1.025] $.
Each pixel corresponds to an element of  the matrix 
such that the matrix index 
of the field $ \phi_{n,k}$ is given by 
$ i= k(2N_t+N_\tau) +n$ with $ k=0,\cdots, N_s-1$ and 
$ n=0,\cdots, 2N_t +N_\tau-1$.
We observe that the real part of $K_\textrm{opt}$ is 
dominated by the diagonal, while in the imaginary 
part one recognizes a banded structure differentiating
between the branches of the contour.
We generally see that non-diagonal values are small: the further away from the diagonal they are, the smaller values are observed.
We also show an inset with a zoom in to a single 
spatial slice, where we note that a structure of the matrix is very similar to that 
observed in 0+1 dimensional calculations \cite{Alvestad:2022abf,Lampl:2023xpb}.

\section{Conclusions}

In summary, we have shown that 
one can exploit the kernel freedom of the CLE
to improve real-time simulations of scalar 
field theories.
We used a dense constant matrix kernel 
in an 1+1 dimensional scalar field theory, and we 
have achieved correct simulations on time contours longer that what is available with other simulation methods.
It is expected that further increasing the temporal extent will lead to an eventual failure as well, investigating the limit is an ongoing project.
The most expensive part of the simulations 
is the training of the kernel, which we optimized 
with machine learning methods. 
We see various possibilities to improve upon the 
current result:
First, we observe that a dense kernel is not needed to achieve convergence, and a sparser kernel
(using e.g. a convolutional matrix) would most likely 
achieve the same benefits while having much reduced costs.
Furthermore, it is expected that a field dependent kernel can be even more beneficial in reducing boundary terms 
at large temporal sizes. Lastly, extension of the 
idea to the more physically relevant 3+1 dimensions
is also underway.

\begin{acknowledgments}
   This research was funded 
in part by the Austrian Science Fund (FWF) via the Principal Investigator Project 
\href{https://doi.org/10.55776/P36875}{P36875}. 
D. A. and A. R. received support from the Research
Council of Norway under the FRIPRO Young Research
Talent grant 286883. The study has benefited from
computing resources provided by UNINETT Sigma2 - the
National Infrastructure for High Performance Computing
and Data Storage in Norway under project NN9578KQCDrtX ”Real-time dynamics of nuclear matter under
extreme conditions”. Some parts of the numerical calculations where performed on GSC, the computing cluster
of the University of Graz.

  \end{acknowledgments}

\bibliographystyle{jhep}
\bibliography{refs}

\providecommand{\noopsort}[1]{}\providecommand{\singleletter}[1]{#1}%

\providecommand{\href}[2]{#2}\begingroup\raggedright\begin{thebibliography}{10}

\bibitem{gattringer_approaches_2016}
C.~Gattringer and K.~Langfeld, \emph{{Approaches to the sign problem in lattice
  field theory}}, \href{https://doi.org/10.1142/S0217751X16430077}{\emph{Int.
  J. Mod. Phys. A} {\bfseries 31} (2016) 1643007}
  [\href{https://arxiv.org/abs/1603.09517}{{\ttfamily 1603.09517}}].

\bibitem{de_forcrand_constraining_2010}
P.~de~Forcrand and O.~Philipsen, \emph{{Constraining the QCD phase diagram by
  tricritical lines at imaginary chemical potential}},
  \href{https://doi.org/10.1103/PhysRevLett.105.152001}{\emph{Phys. Rev. Lett.}
  {\bfseries 105} (2010) 152001}
  [\href{https://arxiv.org/abs/1004.3144}{{\ttfamily 1004.3144}}].

\bibitem{braun_imaginary_2013}
J.~Braun, J.-W.~Chen, J.~Deng, J.E.~Drut, B.~Friman, C.-T.~Ma et~al.,
  \emph{{Imaginary polarization as a way to surmount the sign problem in $Ab$
  $Initio$ calculations of spin-imbalanced Fermi gases}},
  \href{https://doi.org/10.1103/PhysRevLett.110.130404}{\emph{Phys. Rev. Lett.}
  {\bfseries 110} (2013) 130404}
  [\href{https://arxiv.org/abs/1209.3319}{{\ttfamily 1209.3319}}].

\bibitem{braun_zero-temperature_2015}
J.~Braun, J.E.~Drut and D.~Roscher, \emph{{Zero-temperature equation of state
  of mass-imbalanced resonant Fermi gases}},
  \href{https://doi.org/10.1103/PhysRevLett.114.050404}{\emph{Phys. Rev. Lett.}
  {\bfseries 114} (2015) 050404}
  [\href{https://arxiv.org/abs/1407.2924}{{\ttfamily 1407.2924}}].

\bibitem{guenther_qcd_2017}
J.N.~Guenther, R.~Bellwied, S.~Borsanyi, Z.~Fodor, S.D.~Katz, A.~Pasztor
  et~al., \emph{{The QCD equation of state at finite density from analytical
  continuation}},
  \href{https://doi.org/10.1016/j.nuclphysa.2017.05.044}{\emph{Nucl. Phys. A}
  {\bfseries 967} (2017) 720}
  [\href{https://arxiv.org/abs/1607.02493}{{\ttfamily 1607.02493}}].

\bibitem{Klauder:1983nn}
J.R.~Klauder, \emph{{Stochastic Quantisation}},
  \href{https://doi.org/10.1007/978-3-7091-7651-1_8}{\emph{Acta Phys. Austriaca
  Suppl.} {\bfseries 25} (1983) 251}.

\bibitem{Parisi:1984cs}
G.~Parisi, \emph{{On Complex Probabilities}},
  \href{https://doi.org/10.1016/0370-2693(83)90525-7}{\emph{Phys. Lett. B}
  {\bfseries 131} (1983) 393}.

\bibitem{rom_shifted-contour_1997}
N.~Rom, D.~Charutz and D.~Neuhauser, \emph{Shifted-contour auxiliary-field
  {Monte} {Carlo}: circumventing the sign difficulty for electronic-structure
  calculations},
  \href{https://doi.org/10.1016/S0009-2614(97)00370-9}{\emph{Chem. Phys. Lett.}
  {\bfseries 270} (1997) 382}.

\bibitem{cristoforetti_new_2012}
M.~Cristoforetti, F.~Di~Renzo and L.~Scorzato, \emph{{New approach to the sign
  problem in quantum field theories: High density QCD on a Lefschetz thimble}},
  \href{https://doi.org/10.1103/PhysRevD.86.074506}{\emph{Phys. Rev. D}
  {\bfseries 86} (2012) 074506}
  [\href{https://arxiv.org/abs/1205.3996}{{\ttfamily 1205.3996}}].

\bibitem{Alexandru:2020wrj}
A.~Alexandru, G.~Basar, P.F.~Bedaque and N.C.~Warrington, \emph{{Complex paths
  around the sign problem}},
  \href{https://doi.org/10.1103/RevModPhys.94.015006}{\emph{Rev. Mod. Phys.}
  {\bfseries 94} (2022) 015006}
  [\href{https://arxiv.org/abs/2007.05436}{{\ttfamily 2007.05436}}].

\bibitem{Nishimura:2023dky}
J.~Nishimura, K.~Sakai and A.~Yosprakob, \emph{{A new picture of quantum
  tunneling in the real-time path integral from Lefschetz thimble
  calculations}}, \href{https://doi.org/10.1007/JHEP09(2023)110}{\emph{JHEP}
  {\bfseries 09} (2023) 110}
  [\href{https://arxiv.org/abs/2307.11199}{{\ttfamily 2307.11199}}].

\bibitem{bauer_quantum_2022}
C.W.~Bauer et~al., \emph{{Quantum Simulation for High-Energy Physics}},
  \href{https://doi.org/10.1103/PRXQuantum.4.027001}{\emph{PRX Quantum}
  {\bfseries 4} (2023) 027001}
  [\href{https://arxiv.org/abs/2204.03381}{{\ttfamily 2204.03381}}].

\bibitem{Dalzell:2023ywa}
A.M.~Dalzell et~al., \emph{{Quantum algorithms: A survey of applications and
  end-to-end complexities}},
  \href{https://arxiv.org/abs/2310.03011}{{\ttfamily 2310.03011}}.

\bibitem{Alvestad:2023jgl}
D.~Alvestad, A.~Rothkopf and D.~Sexty, \emph{{Lattice real-time simulations
  with learned optimal kernels}},
  \href{https://doi.org/10.1103/PhysRevD.109.L031502}{\emph{Phys. Rev. D}
  {\bfseries 109} (2024) L031502}
  [\href{https://arxiv.org/abs/2310.08053}{{\ttfamily 2310.08053}}].

\bibitem{Alexandru:2017lqr}
A.~Alexandru, G.~Basar, P.F.~Bedaque and G.W.~Ridgway, \emph{{Schwinger-Keldysh
  formalism on the lattice: A faster algorithm and its application to field
  theory}}, \href{https://doi.org/10.1103/PhysRevD.95.114501}{\emph{Phys. Rev.
  D} {\bfseries 95} (2017) 114501}
  [\href{https://arxiv.org/abs/1704.06404}{{\ttfamily 1704.06404}}].

\bibitem{Berges:2006xc}
J.~Berges, S.~Borsanyi, D.~Sexty and I.O.~Stamatescu, \emph{{Lattice
  simulations of real-time quantum fields}},
  \href{https://doi.org/10.1103/PhysRevD.75.045007}{\emph{Phys. Rev. D}
  {\bfseries 75} (2007) 045007}
  [\href{https://arxiv.org/abs/hep-lat/0609058}{{\ttfamily hep-lat/0609058}}].

\bibitem{Berges:2007nr}
J.~Berges and D.~Sexty, \emph{{Real-time gauge theory simulations from
  stochastic quantization with optimized updating}},
  \href{https://doi.org/10.1016/j.nuclphysb.2008.01.018}{\emph{Nucl. Phys. B}
  {\bfseries 799} (2008) 306}
  [\href{https://arxiv.org/abs/0708.0779}{{\ttfamily 0708.0779}}].

\bibitem{Alvestad:2022abf}
D.~Alvestad, R.~Larsen and A.~Rothkopf, \emph{{Towards learning optimized
  kernels for complex Langevin}},
  \href{https://doi.org/10.1007/JHEP04(2023)057}{\emph{JHEP} {\bfseries 04}
  (2023) 057} [\href{https://arxiv.org/abs/2211.15625}{{\ttfamily
  2211.15625}}].

\bibitem{Lampl:2023xpb}
N.M.~Lampl and D.~Sexty, \emph{{Real time evolution of scalar fields with
  kernelled Complex Langevin equation}},
  \href{https://arxiv.org/abs/2309.06103}{{\ttfamily 2309.06103}}.

\bibitem{Okamoto:1988ru}
H.~Okamoto, K.~Okano, L.~Schulke and S.~Tanaka, \emph{{The Role of a Kernel in
  Complex Langevin Systems}},
  \href{https://doi.org/10.1016/0550-3213(89)90526-9}{\emph{Nucl. Phys. B}
  {\bfseries 324} (1989) 684}.

\bibitem{Okano:1991tz}
K.~Okano, L.~Schulke and B.~Zheng, \emph{{Kernel controlled complex Langevin
  simulation: Field dependent kernel}},
  \href{https://doi.org/10.1016/0370-2693(91)91111-8}{\emph{Phys. Lett. B}
  {\bfseries 258} (1991) 421}.

\bibitem{Hansen:2024kjm}
M.W.~Hansen, M.~Mandl, E.~Seiler and D.~Sexty, \emph{{The Role of Integration
  Cycles in Complex Langevin Simulations}},
  \href{https://arxiv.org/abs/2412.17137}{{\ttfamily 2412.17137}}.

\bibitem{boundaryterms1}
M.~Scherzer, E.~Seiler, D.~Sexty and I.-O.~Stamatescu, \emph{{Complex Langevin
  and boundary terms}},
  \href{https://doi.org/10.1103/PhysRevD.99.014512}{\emph{Phys. Rev. D}
  {\bfseries 99} (2019) 014512}
  [\href{https://arxiv.org/abs/1808.05187}{{\ttfamily 1808.05187}}].

\bibitem{boundaryterms2}
M.~Scherzer, E.~Seiler, D.~Sexty and I.O.~Stamatescu, \emph{{Controlling
  Complex Langevin simulations of lattice models by boundary term analysis}},
  \href{https://doi.org/10.1103/PhysRevD.101.014501}{\emph{Phys. Rev. D}
  {\bfseries 101} (2020) 014501}
  [\href{https://arxiv.org/abs/1910.09427}{{\ttfamily 1910.09427}}].

\end{thebibliography}\endgroup



\end{document}